\documentclass{iopstyle}
\usepackage{graphicx}
\usepackage{subcaption}
\usepackage{amsmath}
\usepackage{siunitx}
\usepackage{amssymb}
\usepackage{fancyhdr}
\usepackage{hyperref}
\usepackage[x11names]{xcolor}
\usepackage{tikz}
\usepackage{booktabs}
\usepackage{caption}
\usetikzlibrary{positioning, shapes.geometric, arrows.meta}

\newcommand{\Dp}{D^+}

\newcommand{\Dz}{D^0}
\newcommand{\Dsp}{D_s^+}

\newcommand{\Lcp}{\Lambda_c^+}
\newcommand{\Lcm}{\bar{\Lambda}_c^-}
\newcommand{\ee}{e^+e^-}

\newcommand{\LcptoXenu}{\Lambda_c^+ \to X e^+ \nu_e}

\newcommand{\mupi}{\mu_\pi^2}

\newcommand{\mupiLc}{\mu_\pi^2(\Lambda_c^+)}
\newcommand{\rhoDLc}{\rho_D^3(\Lambda_c^+)}
\newcommand{\chindf}{\chi^2/n_{\mathrm{d.o.f.}}}

\newcommand{\Vcs}{|V_{cs}|}
\newcommand{\Vcd}{|V_{cd}|}

\newcommand{\gev}{\,\text{GeV}}

\newcommand{\avg}[1]{\langle #1 \rangle} 

\begin{document}

\articletype{Paper}

\title{Bayesian Extraction of HQET Parameters from Inclusive Semi-Leptonic Decay of the $\Lcp$ Baryon}

\author{Kangkang Shao$^1$\orcid{0009-0006-4907-1821}, Dong Xiao$^{2,*}$\orcid{0000-0003-4319-1305}}

\affil{$^1$School of Physics, Huazhong University of Science and Technology, Wuhan 430074, China}

\affil{$^2$School of Nuclear Science and Technology, Lanzhou University, Lanzhou 730000, China}

\affil{$^*$Author to whom any correspondence should be addressed.}

\email{$^{*}$xiaod@lzu.edu.cn (corresponding author)}


\begin{abstract}
We extract the non-perturbative Heavy Quark Effective Theory (HQET) parameters from the inclusive semi-leptonic decay $\LcptoXenu$. Unlike charmed mesons produced near threshold, $\Lcp$ baryons produced in $\ee$ annihilation exhibit a complex momentum distribution, making the transformation of the electron energy spectrum from the laboratory frame to the $\Lcp$ rest frame non-trivial. To address this, we develop a novel Bayesian inference method to reconstruct the electron energy moments in the $\Lcp$ rest frame. By performing a global fit of theoretical predictions in the 1S mass scheme to these extracted moments, we determine the HQET parameters $\mupiLc$ and $\rhoDLc$ for the first time using a purely data-driven approach.
\end{abstract}

\section{Introduction}

Recent years have witnessed significant progress in understanding of heavy-quark dynamics and inclusive decays of heavy hadrons. Accurate theoretical predictions for lifetimes and decay distributions of charm and bottom hadrons play a central role in testing the Standard Model and in extracting fundamental parameters such as CKM matrix elements~\cite{Friday:2025gpj,Egner:2024lay,Black:2024bus,Piscopo:2024wpd}. In this context, the nonperturbative inputs associated with the heavy-quark expansion (HQE) / Heavy Quark Effective Theory (HQET), especially higher-dimensional operators, are among the dominant sources of theoretical uncertainty.

In the charmed meson sector (such as \(\Dz\), \(\Dp\), and \(\Dsp\)), recent analyses have successfully extracted HQET matrix elements by employing the 1S mass scheme to ensure perturbative convergence~\cite{Shao:2025vhe}. Extending such success to the baryon sector is of critical importance. A reliable determination of non-perturbative parameters (e.g.\ the kinetic operator \(\mupi\), the Darwin operator \(\rho_{LS}^3\), and others) for charmed baryons directly impacts our understanding of their lifetime hierarchy~\cite{Cheng:2021qpd,Cheng:2023jpz,Dulibic:2023jeu}, and also affects the precision of inclusive determinations of CKM elements such as \(\Vcs\) and \(\Vcd\)~\cite{Gratrex:2022xpm,LHCb:2018nfa,LHCb:2021vll,Shao:2025qwp,LHCb:2019ldj}.

However, applying the same strategy as in the meson sector to baryons — in particular to \(\Lcp\) — faces a unique experimental challenge. In meson experiments (e.g.\ at threshold or tagged-meson setups), the Lorentz boost of the parent meson is small or fixed, which allows a straightforward transformation of measured electron energy spectrum from the lab frame to the meson rest frame. In contrast, \(\Lcp\) baryons produced in generic \(\ee\) annihilation experiments, such as Belle or BESIII emerge with a broad and mixed momentum distributions. Consequently, the observed laboratory-frame electron spectrum is a superposition of many distinct kinematic configurations, making the relation to the \(\Lcp\) rest-frame electron energy is nontrivial, thereby complicating any attempt to extract rest-frame decay distributions (and thus HQET moments) in a model-independent manner.

In this work, we address this challenge by introducing a novel iterative Bayesian inference framework. We treat the transformation from the laboratory-frame to the \(\Lcp\) rest-frame as a probabilistic inverse problem. Rather than relying on fixed-boost approximations or discarding events with large boost uncertainty, our method infers the most probable underlying rest-frame electron energy distribution directly from the full, superimposed laboratory-frame data. We then combine this novel extraction method with theoretical calculations that include power corrections up to \(\mathcal O(1/m_c^3)\) and perturbative corrections up to NNLO. This methodology maximizes the extraction of physical information from limited statistics, allowing for a precise, model-independent determination of the spectral moments and the \(\Lcp\) HQET parameters in the presence of complex kinematic mixed data.

The structure of the paper is as follows: Section 2 presents our Bayesian inference approach to obtain the electronic energy spectrum in $\Lcp$ rest frame and reports the resulting spectrum and electron energy moments. Section 3 discusses the theoretical expressions of the electron energy moments and performs a global fitting of these formulas to data to constrain relevant HQET parameters for $\Lcp$. 
Finally,  Section 4 discusses the summary and prospects for the HQET parameters determined from inclusive semi-leptonic decay of charmed hadrons.

\section{Calculation of Electron Energy Moments in $\Lcp$ Rest Frame}

\subsection{Iterative Bayesian Inference Methodology}

The analysis of the electron energy spectrum in the semi-leptonic decay $\LcptoXenu$ provides crucial insights into the form factors and internal structure of the $\Lcp$ baryon~\cite{PhysRevD.107.052005}. The data were collected at seven distinct center-of-mass energies ($\sqrt{s}$) ranging from 4.6 to 4.7 GeV at BESIII, where the process $\ee \to \Lcp \Lcm$ occurs. The experimentally measured electron momentum distribution in the laboratory frame is a composite spectrum, aggregating data from all energy points.

A direct transformation of this combined lab-frame spectrum to the $\Lcp$ rest frame is unfeasible due to several complicating factors: 
\begin{itemize}
    \item \textbf{Varying Boosts:} The $\Lcp$ baryons are produced with different Lorentz boosts at each of the seven energy points.
    \item \textbf{Varying Polarization:} The longitudinal polarization ($\alpha_0$) of the $\Lcp$ is energy-dependent, affecting its production angle and consequently the kinematics of the daughter electron in the lab frame.
    \item \textbf{Detector Effects:} The measured spectrum is distorted by the detector's finite geometric acceptance and reconstruction efficiency.
\end{itemize}

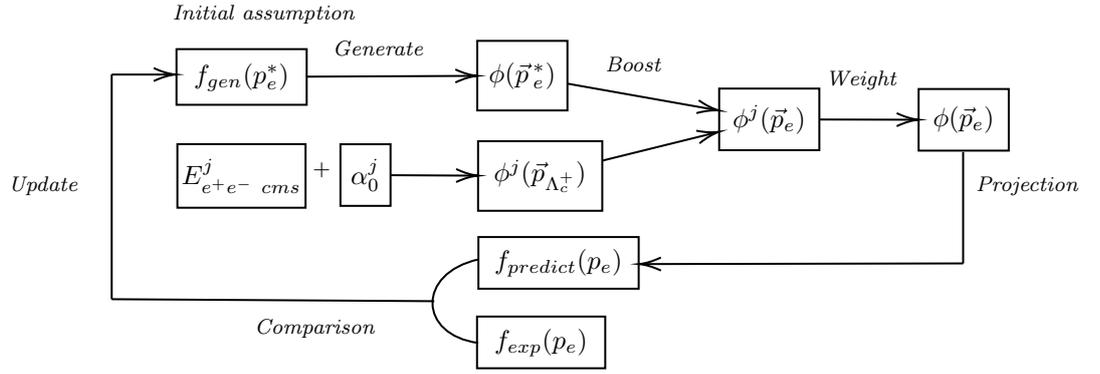
\begin{figure}[h!]
\centering
\tikzset{every picture/.style={line width=0.75pt}}     
\begin{tikzpicture}[x=0.75pt,y=0.75pt,yscale=-1,xscale=1]

\draw    (290.11,215.67) .. controls (259.11,222.67) and (260.11,253.67) .. (290.11,257.67) ;
\draw    (532.11,217.67) -- (372.33,218) ;
\draw [shift={(370.33,218)}, rotate = 359.88] [color={rgb, 255:red, 0; green, 0; blue, 0 }  ][line width=0.75]    (10.93,-3.29) .. controls (6.95,-1.4) and (3.31,-0.3) .. (0,0) .. controls (3.31,0.3) and (6.95,1.4) .. (10.93,3.29)   ;
\draw    (532.11,161.67) -- (532.11,217.67) ;
\draw    (107.11,235.67) -- (268.11,236.67) ;
\draw    (107.11,122.67) -- (107.11,235.67) ;
\draw    (107.11,122.67) -- (137,122.67) ;
\draw [shift={(139,122.67)}, rotate = 180] [color={rgb, 255:red, 0; green, 0; blue, 0 }  ][line width=0.75]    (10.93,-3.29) .. controls (6.95,-1.4) and (3.31,-0.3) .. (0,0) .. controls (3.31,0.3) and (6.95,1.4) .. (10.93,3.29)   ;

\draw (54,172.33) node [anchor=north west][inner sep=0.75pt]  [font=\footnotesize] [align=left] {\textit{Update}};
\draw (537,172.33) node [anchor=north west][inner sep=0.75pt]  [font=\footnotesize] [align=left] {\textit{Projection}};
\draw (177,244.33) node [anchor=north west][inner sep=0.75pt]  [font=\footnotesize] [align=left] {\textit{Comparison}};
\draw (462,119.33) node [anchor=north west][inner sep=0.75pt]  [font=\footnotesize] [align=left] {\textit{Weight}};
\draw (352,112.33) node [anchor=north west][inner sep=0.75pt]  [font=\footnotesize] [align=left] {\textit{Boost}};
\draw (216,104.33) node [anchor=north west][inner sep=0.75pt]  [font=\footnotesize] [align=left] {\textit{Generate}};
\draw (136,86.33) node [anchor=north west][inner sep=0.75pt]  [font=\footnotesize] [align=left] {\textit{Initial assumption}};
\draw (206,165.33) node [anchor=north west][inner sep=0.75pt]  [font=\footnotesize] [align=left] {+};
\draw    (290.16,204.4) -- (370.16,204.4) -- (370.16,230.4) -- (290.16,230.4) -- cycle  ;
\draw (330.16,217.4) node    {$f_{predict} (p_{e} )$};
\draw    (289.46,244.4) -- (353.46,244.4) -- (353.46,270.4) -- (289.46,270.4) -- cycle  ;
\draw (321.46,257.4) node    {$f_{exp} (p_{e} )$};
\draw    (509.89,129.94) -- (554.89,129.94) -- (554.89,160.94) -- (509.89,160.94) -- cycle  ;
\draw (532.39,145.44) node    {$\phi (\vec{p}_{e} )$};
\draw    (221.28,157.72) -- (246.28,157.72) -- (246.28,188.72) -- (221.28,188.72) -- cycle  ;
\draw (233.78,173.22) node    {$\alpha _{0}^{j}$};
\draw    (139.88,157.66) -- (203.88,157.66) -- (203.88,189.66) -- (139.88,189.66) -- cycle  ;
\draw (171.88,173.66) node    {$E_{e^{+} e^{-} \ cms}^{j}$};
\draw    (289.98,156.2) -- (351.98,156.2) -- (351.98,191.2) -- (289.98,191.2) -- cycle  ;
\draw (320.98,173.7) node    {$\phi ^{j} (\vec{p}_{\Lambda _{c}^{+}} )$};
\draw    (410.33,129.72) -- (460.33,129.72) -- (460.33,160.72) -- (410.33,160.72) -- cycle  ;
\draw (435.33,145.22) node    {$\phi ^{j} (\vec{p}_{e} )$};
\draw    (289.59,105.72) -- (334.59,105.72) -- (334.59,140.72) -- (289.59,140.72) -- cycle  ;
\draw (312.09,123.22) node    {$\phi (\vec{p}\,^{*}_{e})$};
\draw    (139.44,109.88) -- (204.44,109.88) -- (204.44,137.88) -- (139.44,137.88) -- cycle  ;
\draw (171.94,123.88) node    {$f_{gen} (p_{e}^{*} )$};
\draw    (460.33,145.28) -- (507.89,145.39) ;
\draw [shift={(509.89,145.39)}, rotate = 180.13] [color={rgb, 255:red, 0; green, 0; blue, 0 }  ][line width=0.75]    (10.93,-3.29) .. controls (6.95,-1.4) and (3.31,-0.3) .. (0,0) .. controls (3.31,0.3) and (6.95,1.4) .. (10.93,3.29)   ;
\draw    (246.28,173.29) -- (287.98,173.52) ;
\draw [shift={(289.98,173.53)}, rotate = 180.31] [color={rgb, 255:red, 0; green, 0; blue, 0 }  ][line width=0.75]    (10.93,-3.29) .. controls (6.95,-1.4) and (3.31,-0.3) .. (0,0) .. controls (3.31,0.3) and (6.95,1.4) .. (10.93,3.29)   ;
\draw    (351.98,165.98) -- (408.39,151.93) ;
\draw [shift={(410.33,151.45)}, rotate = 166.02] [color={rgb, 255:red, 0; green, 0; blue, 0 }  ][line width=0.75]    (10.93,-3.29) .. controls (6.95,-1.4) and (3.31,-0.3) .. (0,0) .. controls (3.31,0.3) and (6.95,1.4) .. (10.93,3.29)   ;
\draw    (334.59,127.24) -- (408.36,140.41) ;
\draw [shift={(410.33,140.76)}, rotate = 190.12] [color={rgb, 255:red, 0; green, 0; blue, 0 }  ][line width=0.75]    (10.93,-3.29) .. controls (6.95,-1.4) and (3.31,-0.3) .. (0,0) .. controls (3.31,0.3) and (6.95,1.4) .. (10.93,3.29)   ;
\draw    (204.44,123.73) -- (287.59,123.34) ;
\draw [shift={(289.59,123.33)}, rotate = 179.73] [color={rgb, 255:red, 0; green, 0; blue, 0 }  ][line width=0.75]    (10.93,-3.29) .. controls (6.95,-1.4) and (3.31,-0.3) .. (0,0) .. controls (3.31,0.3) and (6.95,1.4) .. (10.93,3.29)   ;

\end{tikzpicture}
\caption{A flowchart illustrating the iterative Bayesian inference method. The process starts with an arbitrary assumption for the electron momentum probability density function (PDF) in the $\Lcp$ rest frame, which is refined by comparing the resulting simulated lab-frame PDF with experimental data. Here $f$ is the one dimensional normalized PDF of electron momentum, while $\phi$ denotes the PDF in the three momentum vector space. The $p$ and $\vec{p}$ represent the variables in the laboratory frame, whereas $p^*$ and $\vec{p}\,^*$ stand for the variables in the $\Lcp$ rest frame.}
\label{fig:flowchart}
\end{figure}

To address these challenges, we employ an iterative Bayesian inference technique, summarized in Figure~\ref{fig:flowchart}. This method uses a Monte Carlo (MC) simulation to model the entire process from a hypothesized true electron spectrum in the $\Lcp$ rest frame to the measured electron spectrum in lab frame, and iteratively refines the hypothesis until the simulation matches the data.

The core of the analysis is an iterative procedure based on the principles of Bayes' theorem, as pioneered by D'Agostini~\cite{dagostini2010}.

\subsection{Monte Carlo Simulation}
A detailed MC simulation is the engine of the Bayesian procedure. For each iteration, a large sample of events is generated following these steps:
\begin{enumerate}
    \item \textbf{Energy Point Sampling:} An energy point ($\sqrt{s}$) is randomly selected, weighted by the measured single-tag yields of $\Lcm$ at each point, ensuring the correct mixture of production conditions.
    \item \textbf{$\Lcp$ Generation (Lab Frame):} A four-momentum for the parent $\Lcp$ is generated in the laboratory frame. This step correctly models the production angular distribution, which is proportional to $1 + \alpha_0(\sqrt{s}) \cos^2\theta$, using the experimentally measured energy-dependent polarization parameter $\alpha_0(\sqrt{s})$. A small transverse boost of the $e^+e^-$ system is also included.
    \item \textbf{Electron Generation ($\Lcp$ Rest Frame):} An electron momentum, $p^*_e$, is randomly sampled from the current hypothesis for the ``origin" spectrum, $f_{gen} (p^*_e)$. A four-momentum is then constructed assuming isotropic decay in the $\Lcp$ rest frame, distributed as $\phi (\vec{p}\,^{*}_{e})$ .
    \item \textbf{Transformation to Lab Frame:} The electron's four-momentum is then Lorentz-boosted from the $\Lcp$ rest frame to the laboratory frame.
    \item \textbf{Detector Emulation:} A cut on the electron's lab-frame polar angle, $|\cos\theta_{\text{lab}}| < 0.93$, is applied to emulate the geometric acceptance of the detector.
\end{enumerate}
This process produces a predicted lab-frame spectrum, $f_{predict} (p_e)$, and a migration matrix $\mathcal{M}$, which maps the relationship between the generated electron momentum in $\Lcp$ rest frame and observed electron momentum in lab frame.

\subsection{The Iterative Procedure}

The iteration procedure is performed as follows. For each iteration $k$:

\begin{enumerate}
    \item Generate a Monte Carlo sample based on the current hypothesis of electron momentum distribution in $\Lcp$ rest frame, $f^{k}_{gen}(p\,^{*}_{e})$, to produce a predicted lab-frame spectrum, $f^k_{predict} (p_e)$, and a migration matrix $\mathcal{M}^k$. The element $\mathcal{M}^k_{ij}$ is defined as the number of electrons observed in the $j$-th momentum bin in the lab frame that originated from the $i$-th momentum bin in the $\Lcp$ rest frame.

    \item Calculate the efficiency, $\varepsilon^k_i$, of observing an electron generated within $i$-th momentum bin of the $\Lcp$ rest frame.
    \begin{equation}
        \varepsilon^k_i = \frac{\sum_{j} \mathcal{M}^k_{ji}}{N^k_{gen,i}}
    \end{equation}

    \item Estimate the Bayesian probability, $P^k(i|j)$, that an electron observed in $j$-th lab-frame momentum bin originated from the $i$-th momentum bin in the $\Lcp$ rest frame.
    \begin{equation}
        P^k(i|j) = \frac{\mathcal{M}^k_{ij}}{\sum_{i} \mathcal{M}^k_{ij}}
    \end{equation}
    
    \item Update the hypothesis for the electron momentum distribution in the $\Lcp$ rest frame for the next iteration based on the experimental measured distribution in the lab frame.
    \begin{equation}
        f^{k+1}_{gen,i}(p\,^{*}_{e}) = \frac{1}{\varepsilon^k_i} \cdot \sum_{j}{f_{exp,j}(p_e) \cdot P^k(i|j)}
    \end{equation}
    where $f_{exp,j}(p_e)$ is the content of the $j$-th bin of the experimental data.
\end{enumerate}

For each iteration, we generate 10,000,000 events containing the information of electron's four-momentum vector in the $\Lcp$ rest frame and $\Lcp$'s four-momentum vector in the lab-frame weighted at seven distinctive CMS energies. Then repeat steps from 1 to 4 until the $\chindf$ between the predicted $f_{predict}(p_e)$ and measured $f_{exp}(p_e)$ reaches the value closest to 1.

\subsection{Low-Momentum Theoretical Constraint}
The experimental data contain no information below $p_e = 0.2$ GeV/c. To prevent unphysical behavior, the unfolded spectrum in the region $p_e^* \in [0, 0.2]$ GeV/c is constrained to follow the theoretical form $f(p_e^*) \propto (p_e^*)^2(1+bp_e^*)(1-p_e^*)$. The parameters of this function are determined at each iteration by enforcing continuity in both value and first-derivative (C$^1$) at the boundary point ($p_e^*=0.2$ GeV/c), ensuring a smooth transition between the theory-driven and data-driven regions of the spectrum.

\subsection{Electron Momentum Distribution}
\begin{figure}[htbp]
    \centering
    \begin{subfigure}[b]{0.48\textwidth}
        \centering
        \includegraphics[width=\textwidth]{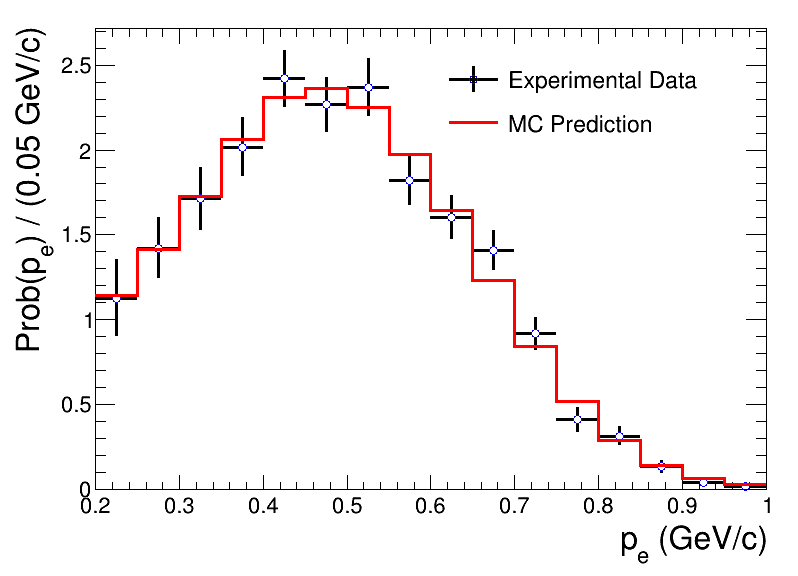}
        \caption{Comparison between simulated and measured PDF of $p_e$ in lab-frame, with $\chindf=0.74$.}
        \label{fig:result_pe_comparison}
    \end{subfigure}
    \hfill 
    \begin{subfigure}[b]{0.48\textwidth}
        \centering
        \includegraphics[width=\textwidth]{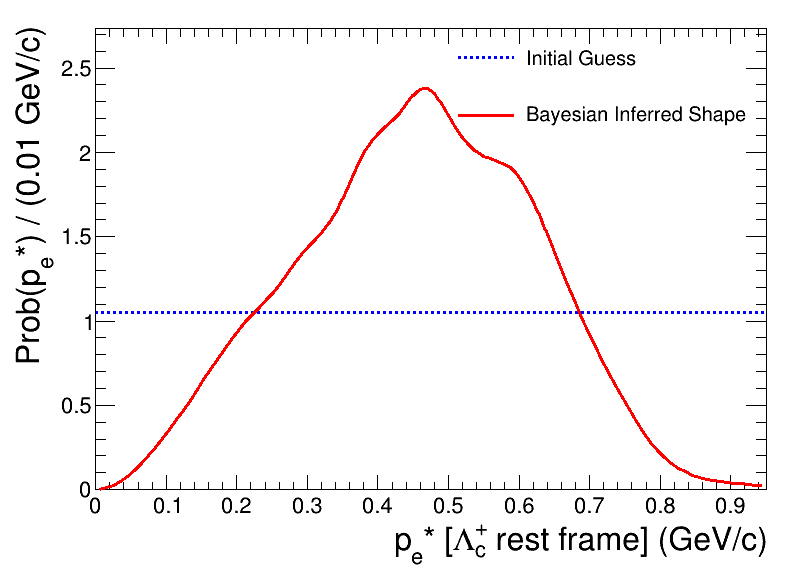}
        \caption{Bayesian inferred PDF of $p^*_e$ in $\Lcp$ rest frame, with the initial assumption of a flat distribution.}
        \label{fig:result_pe_Lcrest}
    \end{subfigure}
    \caption{Result of the electron momentum distribution of $\LcptoXenu$.}
    \label{fig:result_pe}
\end{figure}

The iterative procedure is found to converge after 4 iterations, at which point the $\chindf$ between the final simulated spectrum and the experimental data reaches a stable minimum of 0.74. The final comparison is shown in Figure~\ref{fig:result_pe_comparison}. The Bayesian inferred electron momentum spectrum in the $\Lcp$ rest frame is shown in Figure~\ref{fig:result_pe_Lcrest}.

\subsection{Nominal Results of Electron Energy Moments}
The first four terms of electron energy moments of the decay of $\LcptoXenu$ are defined by integration over the energy spectrum $E_e^*$ in the $\Lcp$ rest frame.

\begin{equation}
\begin{split}
    \avg{E_e} & = \frac{1}{\Gamma_{\Lcp}}\int \frac{d\Gamma}{dE_e^*} E_e^* dE_e^* \\
    \avg{E_e^2}_\mathrm{center} & = \frac{1}{\Gamma_{\Lcp}}\int \frac{d\Gamma}{dE_e^*} (E_e^*-\avg{E})^2 dE_e^* \\
    \avg{E_e^3}_\mathrm{center} & = \frac{1}{\Gamma_{\Lcp}}\int \frac{d\Gamma}{dE_e^*} (E_e^*-\avg{E})^3 dE_e^* \\
    \avg{E_e^4}_\mathrm{center} & = \frac{1}{\Gamma_{\Lcp}}\int \frac{d\Gamma}{dE_e^*} (E_e^*-\avg{E})^4 dE_e^* \\
\end{split}
\end{equation}

Upon integration, we obtain the following values for the electron energy moments: $\avg{E_e} = 0.455\,\gev$, $\avg{E_e^2}_\mathrm{center} = 2.75\times10^{-2}\,\gev^2$, $\avg{E_e^3}_\mathrm{center} = -3.67\times10^{-4}\,\gev^3$, $\avg{E_e^4}_\mathrm{center} = 1.89\times10^{-3}\,\gev^4$.

\subsection{Statistical Uncertainty}
The statistical uncertainty on the inferred electron energy moments, arising from the finite statistics of the input data, is determined using a ``bootstrap" method. 
\begin{enumerate}
    \item A large number ($N_{\mathrm{toys}}=10000$) of ``toy" experimental spectra are generated by fluctuating the content of each bin of the original data according to a Gaussian distribution defined by its central value and error.
    \item The entire iteration procedure is performed for each toy spectrum.
    \item The statistical error on each final moment is taken as the root mean square (RMS) of the distribution of that moment over all toy experiments.
    \item The full $4 \times 4$ statistical covariance matrix between the moments is also computed from this ensemble.
\end{enumerate}

\begin{table}[h!]
    \centering
    \caption{The statistical covariance matrix for the four electron energy moments, as determined by the bootstrap method.}
    \label{tab:stat_cov_matrix}
    \sisetup{
        scientific-notation = true,
        table-format = +1.3e+2,
        round-mode = places,
        round-precision = 3
    }
    \begin{tabular}{c S[table-format = +1.4e+2]
                      S S S}
        \toprule
        Stat. Cov. & \multicolumn{1}{c}{$\avg{E_e}/\gev$} & \multicolumn{1}{c}{$\avg{E_e^2}_\mathrm{center}/\gev^2$} & \multicolumn{1}{c}{$\avg{E_e^3}_\mathrm{center}/\gev^3$} & \multicolumn{1}{c}{$\avg{E_e^4}_\mathrm{center}/\gev^4$} \\
        \midrule
        $\avg{E_e}/\gev$ &  2.472e-04 & -5.315e-05 &  4.762e-06 & -6.735e-06 \\
        $\avg{E_e^2}_\mathrm{center}/\gev^2$ & -5.315e-05 &  1.314e-05 & -1.206e-06 &  1.634e-06 \\
        $\avg{E_e^3}_\mathrm{center}/\gev^3$ &  4.762e-06 & -1.206e-06 &  1.926e-07 & -1.404e-07 \\
        $\avg{E_e^4}_\mathrm{center}/\gev^4$ & -6.735e-06 &  1.634e-06 & -1.404e-07 &  2.011e-07 \\
        \bottomrule
    \end{tabular}
\end{table}

\subsection{Systematic Uncertainties}
The dominant systematic uncertainties in this Bayesian inference procedure arise from three primary sources: the choice of the number of iterations, the initial shape assumed for the distribution in $\Lcp$ rest frame, and the experimental uncertainties on the polarization parameters ($\alpha_0$). To robustly estimate these uncertainties and account for potential correlations between them, we employ a grid scan approach.

We define a set of discrete variations for each of the three main sources:
\begin{enumerate}
    \item \textbf{Number of Iterations:} The number of Bayesian iterations is varied across three choices: \{3, 4, 5\}, centered around our nominal choice of 4.
    \item \textbf{Initial Shape:} Three distinct functional forms are used for the initial guess of the true spectrum: a flat distribution (nominal), a polynomial function, and a sine function.
    \item \textbf{Polarization Parameter ($\alpha_0$):} The set of $\alpha_0(\sqrt{s})$ parameters is varied according to its total experimental uncertainty: \{Nominal - $1\sigma$, Nominal, Nominal + $1\sigma$\}.
\end{enumerate}

The full procedure is performed for all $3 \times 3 \times 3 = 27$ combinations of these discrete choices. This ensemble of 27 output moment sets, $\{ \avg{E_e}, \avg{E_e^2}_\mathrm{center}, \avg{E_e^3}_\mathrm{center}, \avg{E_e^4}_\mathrm{center} \}$, represents the space of the systematic model variations.

The total systematic uncertainty for each moment is defined as the RMS of this distribution. Furthermore, this method allows for the calculation of the full $4 \times 4$ systematic covariance matrix, which captures the correlations in how the moments shift together as the analysis assumptions are varied.

\begin{table}[h!]
    \centering
    \caption{The systematic covariance matrix for the four electron energy moments, as determined by the grid scan method.}
    \label{tab:syst_cov_matrix}
    \sisetup{
        scientific-notation = true,
        table-format = +1.3e+2,
        round-mode = places,
        round-precision = 3
    }
    \begin{tabular}{c S[table-format = +1.4e+2]
                      S S S}
        \toprule
        Sys. Cov. & \multicolumn{1}{c}{$\avg{E_e}/\gev$} & \multicolumn{1}{c}{$\avg{E_e^2}_\mathrm{center}/\gev^2$} & \multicolumn{1}{c}{$\avg{E_e^3}_\mathrm{center}/\gev^3$} & \multicolumn{1}{c}{$\avg{E_e^4}_\mathrm{center}/\gev^4$} \\
        \midrule
        $\avg{E_e}/\gev$ &  7.031e-06 & -1.716e-06 &  1.914e-07 & -1.791e-07 \\
        $\avg{E_e^2}_\mathrm{center}/\gev^2$ & -1.716e-06 &  4.462e-07 & -3.776e-08 &  4.722e-08 \\
        $\avg{E_e^3}_\mathrm{center}/\gev^3$ &  1.914e-07 & -3.776e-08 &  8.364e-09 & -1.412e-09 \\
        $\avg{E_e^4}_\mathrm{center}/\gev^4$ & -1.791e-07 &  4.722e-08 & -1.412e-09 &  6.578e-09 \\
        \bottomrule
    \end{tabular}
\end{table}

\subsection{Final Result of Electron Energy Moment}
After considering the statistical and systematic uncertainties, we obtain the final result of the electron energy moment of $\LcptoXenu$ based on the Bayesian inference approach.

\begin{figure}[htbp]
    \centering
    \includegraphics[width=0.55\textwidth]{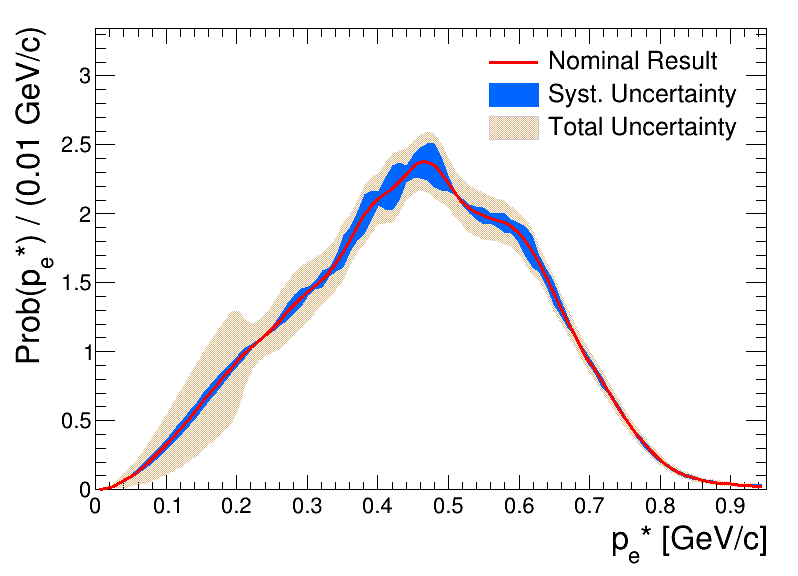}
    \caption{Final result of the electron momentum distribution of $\LcptoXenu$ in the $\Lcp$ rest frame.}
    \label{fig:result_pe_final}
\end{figure}

The Bayesian inferred electron momentum distribution in $\Lcp$ rest frame with error band is shown in Figure~\ref{fig:result_pe_final}.

The integrated results of the first four electron energy moments of $\LcptoXenu$ with statistical and systematic uncertainty are listed in Table~\ref{tab:result_moment}.

\begin{table}[!htb]
    \centering
    \caption{The first four electron energy moments of $\LcptoXenu$. The first uncertainty is statistical, and the subsequent one is the systematic uncertainties.}
    \label{tab:result_moment}
    \begin{tabular}{cc}
        \hline
        \hline
        Energy Moment & Value \\
        \hline
        $\avg{E_e} [\gev]$                      & $ (4.55\pm0.16\pm0.03)\times10^{-1} $ \\
        $\avg{E_e^2}_\mathrm{center} [\gev^2]$  & $ (2.75\pm0.37\pm0.07)\times10^{-2} $ \\
        $\avg{E_e^3}_\mathrm{center} [\gev^3]$  & $ (-3.67\pm4.39\pm0.91)\times10^{-4} $ \\
        $\avg{E_e^4}_\mathrm{center} [\gev^4]$  & $ (1.89\pm0.45\pm0.08)\times10^{-3} $ \\
        \hline
        \hline
    \end{tabular}
\end{table}

\section{Phenomenological Analysis of Inclusive $\Lcp$ Decays}
\subsection{Theoretical formulas}
As in the meson case, the inclusive decay of $\Lcp$ is described within the framework of the heavy-quark expansion, where the short-distance coefficients are computed perturbatively and the resulting expressions are performed as a double expansion in $\Lambda_{\mathrm{QCD}}/m_c$ and $\alpha_s$, with the strange-quark mass treated as a quantity of order $\Lambda_{\mathrm{QCD}}$. For our theoretical formulas, the power corrections are included up to order $(\Lambda_{\mathrm{QCD}}/m_c)^3$. For the leading-power contributions, we incorporate $\alpha_s$ corrections up to NNLO, whereas for the higher-power terms only the leading-order contributions are taken into account. In comparison to the D-meson case, the heavy-quark expansion for $\Lcp$ involves a different pattern of non-perturbative matrix elements: the  chromomagnetic operator $\mu_{G}^2(\Lcp)$ and spin–orbit operators $\rho_{LS}^{3}(\Lcp)$ contributions both vanish in the $\Lcp$,  due to the spin-zero configuration of the light diquark~\cite{Bernlochner:2024vhg}. As a result,  the Darwin term, together with several four-quark operators, play a more prominent role.
We adopt the 1S mass scheme in our analysis, as it provides notably better convergence in the perturbative expansion for charm inclusive decays~\cite{Shao:2025vhe}.

Within the pole mass scheme, the decay widths can be expressed as
{
\footnotesize
\begin{align}\label{eq:decaywidth}
\Gamma_{\Lcp}=\sum_{q=d,s}\hat{\Gamma}_0\left|V_{cq}\right|^2 m_c^5 \Big\{ 1 &+\frac{\alpha_s}{\pi} {2\over3}\left(\frac{25}{4}- {\pi^2}\right) + \frac{\alpha_s^2}{\pi^2}\left[  {\beta_0\over 4} \frac{2}{3} \left(\frac{25}{4}-\pi ^2\right)\log \left(\frac{\mu^2}{m_c^2}\right) +2.14690 n_l-29.88311 \right]  \nonumber \\ 
  & -8\rho\delta_{sq} -\frac{1}{2}\frac{\mu_{\pi}^{2}(\Lcp)}{m_{c}^2} +(\frac{20}{3}+8\log(\frac{\mu^2}{m_{c}^2})) \frac{\rho_D^3\left(\Lcp\right)}{m_c^3}+\frac{\tau_0(\Lcp)}{m_c^3}+ ...\Big\},
\end{align}
}
where $\hat{\Gamma}_0 = G_F^2 /(192\pi^3)$,$V_{cq}$ is the CKM matrix element, and $\rho = m_s^2/m_c^2$. The quantities $\tau_{\Lcp} =128 \pi^2\left(T_1\left(\Lcp\right)-T_2\left(\Lcp\right)\right) $ denote contributions from the four-quark matrix elements for $\Lcp$~\cite{Fael:2019umf}. The QCD coefficient is $\beta_0 = 11 - 2n_f/3$, and we use $n_f = 4$. For the definitions of the local operator matrix elements, we follow the convention of Ref.~\cite{Fael:2019umf}.

 For the initial four raw moments, the theoretical expressions are shown as follows,
 {
 \footnotesize
 \begin{align}
\langle E_{e}\rangle&=\frac{\hat{\Gamma}_0}{    \Gamma_{\Lcp}}\sum_{q=d,s} \left|V_{cq}\right|^2 m_{c}^6 \left[\frac{3}{10} + \frac{\alpha_{s}}{\pi} a_1^{(1)} + {\alpha_s^2\over \pi^2}a_1^{(2)} -3\rho\delta_{sq} +(\frac{77}{15}+4\log(\frac{\mu^2}{m_{c}^2})) \frac{\rho_D^3\left(\Lcp\right)}{m_c^3}+\frac{\tau_0(\Lcp)}{2 m_c^3}+ ...\right], \nonumber\\
\avg{E^2_e} &= \frac{\hat{\Gamma}_0}{\Gamma_{\Lcp}} \sum_{q=d,s}\left|V_{cq}\right|^2 m_c^7 \left[ \frac{1}{10} + \frac{\alpha_{s}}{\pi} a_2^{(1)}  + {\alpha_s^2\over \pi^2}a_2^{(2)}  - \frac{6}{5} \rho\delta_{sq} + \frac{1}{12}\frac{\mu_{\pi}^2(\Lcp)}{ m_c^2} +(\frac{181}{60}+2\log(\frac{\mu^2}{m_{c}^2})) \frac{\rho_D^3\left(\Lcp\right)}{m_c^3} \right.\nonumber \\
&\qquad\qquad\qquad\qquad\qquad \left. +\frac{\tau_0(\Lcp)}{4 m_c^3}+ ... \right], \nonumber\\
\avg{E^3_e} &= \frac{\hat{\Gamma}_0}{\Gamma_{\Lcp}}\sum_{q=d,s}\left|V_{cq}\right|^2 m_c^8\left[\frac{1}{28} + \frac{\alpha_{s}}{\pi} a_3^{(1)}  + {\alpha_s^2\over \pi^2}a_3^{(2)}  -\frac{1}{2} \rho\delta_{sq} +\frac{1}{14} \frac{\mu_\pi^2\left(\Lcp\right)}{ m_c^2}+(\frac{233}{140}+\log(\frac{\mu^2}{m_{c}^2})) \frac{\rho_D^3\left(\Lcp\right)}{m_c^3}\right. \nonumber\\
&\qquad\qquad\qquad\qquad\qquad \left.+\frac{\tau_0(\Lcp)}{8 m_c^3}+ ... \right],\nonumber\\
\avg{E^4_e} &= \frac{\hat{\Gamma}_0}{\Gamma_{\Lcp}}\sum_{q=d,s}\left|V_{cq}\right|^2 m_c^9\left[\frac{3}{224} + \frac{\alpha_{s}}{\pi} a_4^{(1)}  + {\alpha_s^2\over \pi^2}a_4^{(2)}  -\frac{3}{14} \rho\delta_{sq} +\frac{3}{64} \frac{\mu_\pi^2\left(\Lcp\right)}{ m_c^2}+(\frac{1989}{2240}+\frac{1}{2}\log(\frac{\mu^2}{m_{c}^2}))\frac{\rho_D^3\left(\Lcp\right)}{m_c^3}\right. \nonumber\\
&\qquad\qquad\qquad\qquad\qquad \left.+\frac{\tau_0(\Lcp)}{16 m_c^3}+... \right],\label{eq:Emoments}
\end{align}
}
where the coefficients follow the conventions of Ref.~\cite{Shao:2025vhe}. They are highly sensitive to the charm quark mass $m_c$. Although HQET is often formulated in the pole-mass scheme, this choice is not suitable for precision analyses, since the decay width suffers from a renormalon ambiguity. In the $\overline{\rm MS}$ scheme, the perturbative expansion converges rather slowly~\cite{Gray:1990yh,Broadhurst:1991fy,Fleischer:1998dw,Melnikov:2000qh}. For the kinetic mass scheme, the higher-order corrections scale as $(\alpha_s/\pi)\mu^{n}/m_{c}^{n}$, and the choice of the cutoff scale $\mu$ introduces an additional subtlety~\cite{Fael:2019umf,Boushmelev:2023kmf}.

In contrast, the 1S mass scheme avoids the renormalon ambiguity of the pole mass and shows a much better perturbative behavior than the $\overline{\rm MS}$ and kinetic schemes. Its perturbative series is known to be more stable, and the residual scale dependence is substantially reduced\cite{Hoang:1998ng,Hoang:1998hm, Hoang:1999zc}. For these reasons, the 1S mass scheme provides the most reliable input for the inclusive $\Lcp$ analysis and is adopted throughout this scheme.

The pole mass $m_c$ is related to the 1S mass $m_{c,\mathrm{1S}}$ through~\cite{Shao:2025vhe}, 
{
\footnotesize
\begin{equation}\label{eq:1smass}
m_c =m_{c,\mathrm{1S}}+ m_{c,\mathrm{1S}} \frac{ \alpha_s(\mu)^2 C_F^2 }{8}\left\{ 1 + {\alpha_s\over\pi} \left[ \left( - \log\left(\alpha_s(\mu) m_{c,\mathrm{1S}} C_F/\mu\right)+  {11\over6}\right)\beta_0 -4 + {\pi\over 8}C_F \alpha_s \right] + ... \right\}  .
\end{equation}
}
Specifically, we replace the pole mass in Eqs.\eqref{eq:decaywidth} and \eqref{eq:Emoments} with the 1S-mass relation in Eq.\eqref{eq:1smass}, and then expand results up to $O(\alpha_s^2)$. In fact, it is worth noting that the NLO correction constitutes the leading order contribution, since the $\varepsilon$-expansion starts at NLO~\cite{Hoang:1998ng}.

In addition, to suppress the correlations among the experimental data points, we use the electron-energy central moments rather than the raw moments. The experimental data inputs $$\{\Gamma_{\Lcp}, \left\langle E_e\right\rangle,\left\langle E_e^2\right\rangle_{\rm{center}},\left\langle E_e^3\right\rangle_{\rm{center}},\left\langle E_e^4\right\rangle_{\rm{center}}/\left\langle E_e^2\right\rangle_{\rm{center}}\}$$ are then employed in the global fit.
\subsection{Towards an extraction of the kinetic energy and Darwin terms}
To obtain the experimental data points needed to constrain the basic HQET parameters of the $\Lcp$, we develop and employ a Bayesian inference method to reconstruct the electron-energy moments in the $\Lcp$ rest frame, thereby providing precise experimental inputs for the fit. These reconstructed moments serve as the experimental inputs for our global analysis. The corresponding theoretical expressions used in this fit are given in Eqs~\eqref{eq:decaywidth} and~\eqref{eq:Emoments}. 

The numerical inputs required for our analysis are taken from well-established lattice and perturbative QCD determinations. For the strange-quark mass, we employ the $2\!+\!1\!+\!1$ FLAG average
    $\overline{m}_s(2~\mathrm{GeV}) = (93.44 \pm 0.68)\,\mathrm{MeV}$, as reported in Ref.~\cite{FlavourLatticeAveragingGroup:2019iem}.
The strong coupling at the charm scale is fixed to $\alpha_s(\overline{m}_c)=0.387$, following the determination of Ref.~\cite{King:2021xqp} based on the \texttt{RunDec} evolution package~\cite{Herren:2017osy}.
The scale-dependent quantities $\alpha_s(\mu)$ and $\overline{m}_c(\mu)$ are evaluated for renormalization scales in the interval $
    1~\mathrm{GeV} \le \mu \le 2\,\overline{m}_c(\overline{m}_c)$.
For the remaining Standard Model parameters entering the decay-width and moment calculations, we adopt the PDG~2024 values~\cite{ParticleDataGroup:2024cfk}: $G_F = 1.1663788\times 10^{-5}$, $|V_{cs}| = 0.975 \pm 0.006$, $|V_{cd}| = 0.221 \pm 0.004$.

Our $\chi^2$ function is defined as follows
\begin{equation}
\chi^2(\boldsymbol{\theta})=(\boldsymbol{y}-\boldsymbol{\eta}(\boldsymbol{\theta}))^{\mathrm{T}} \boldsymbol{V}^{-1}(\boldsymbol{y}-\boldsymbol{\eta}(\boldsymbol{\theta})),
\end{equation}
where $\boldsymbol{\theta}=\{\mu_{\pi}^2(\Lcp),\rho^{3}_{D}(\Lcp),\tau_{\rm{WA}}(\Lcp)\}$, and $\tau_{\rm{WA}}(\Lcp)\}=|V_{cs}|^2 \tau_{0,\rm{non-valance}}+|V_{cd}|^2 \tau_{0,\rm{valance}}$.

The results\footnote{Within the pseudo-data based robustness tests we performed, we found no evidence for over-fitting} of the global fit are presented in Table~\ref{tab: globalFitting}. 
\begin{table}[!htb]
\centering
\caption{The $\chi^2$ fitting results in the $\mathrm{1S}$ mass scheme.  The $\chi^2$/d.o.f. in the fit, along with the central values and uncertainties for the HQET parameters, are displayed. The first uncertainty arises from experiment data, the second from the evolution of renormalization scale $\mu$ from 1 to 2.54 $\rm{GeV}$, and the last uncertainty comes from unknown higher order power corrections.}
\label{tab: globalFitting}
\begin{tabular}{l c c c c}
\hline
\scriptsize{$\mu_{\pi}^2[10^{-1}\gev^2]$} & \scriptsize{$\rho^{3}_{D}[10^{-4}\gev^3]$} & \scriptsize{$\tau_{\rm{WA}}[10^{-1}\gev^3]$} & \scriptsize{$\chi^2/\rm d.o.f.$} & \scriptsize{$N_{data}$ v.s. $N_{param}$} \\
\hline
\scriptsize{$1.33\pm0.18\pm0.02\pm0.40$} & \scriptsize{$-2.95\pm2.59\pm0.36\pm1.18$} & \scriptsize{$-2.69\pm0.33\pm0.12\pm0.81$} & \scriptsize{0.53} & \scriptsize{5 v.s. 3} \\
\hline
\end{tabular}
\end{table}

Based on the model-independent method, the fit results obtained in the 1S mass scheme  are summarized in Table~\ref{tab: globalFitting}. As the current analysis is limited by the number of available experimental and theoretical inputs, we do not include additional uncertainties associated with unknown higher-order power corrections. The fitted values of $\mu_\pi^2(\Lcp)$ and $\rho_D^{3}(\Lcp)$ are consistent with those extracted in the D-meson sector\cite{Shao:2025vhe}. Conversely, the weak-annihilation parameter $\tau_{\rm WA}(\Lcp)$ exhibits a significantly larger magnitude, reflecting the well-known difference between baryons and mesons in the dimension-6 four-quark contributions~\cite{King:2021xqp,King:2021jsq,Kirk:2017juj}. Moreover, in order to include the possible effects of omitted higher-order power corrections, we introduce a $30\%$ systematic uncertainty in the fitting results. The results presented in Table~\ref{tab: globalFitting} differ from those obtained using the improved bag-model~\cite{Cheng:2023jpz} and the wave-function approach~\cite{Gratrex:2022xpm}.  This discrepancy warrants further investigation in future work.

\section{Conclusion}
In this work, we have introduced a novel Bayesian inference method to reconstruct the electron energy moments in the $\Lcp$ rest-frame. This methodology represents an advancement in handling inclusive semi-leptonic decays where the parent particle's momentum is not fixed. Beyond the specific case of $\Lcp$, this strategy also inspire thoughts for analyzing other inclusive decays at $B$-factories and future colliders. By inferring the rest-frame spectrum from a single, kinematically complex laboratory distribution, we demonstrate a robust alternative to traditional tagging or approximate boost techniques.

By combining these reconstructed moments with a global fit of theoretical predictions in the 1S mass scheme, we have for the first time determined the non-perturbative heavy-quark parameters $\mu_\pi^2(\Lcp)$ and $\rho_D^3(\Lcp)$ directly from data. Our analysis demonstrates that a data-driven approach, which avoids model dependence, is viable and yields stable values for the HQET parameters. These results lay a new foundation for inclusive analyses of charm-baryon decays, bridging semi-leptonic observables with HQET parameter extraction.

%
%

\ack{
We wish to thank Matteo Fael and Wen-Jie Song for enlightening discussions on calculations of electronic energy spectrum, and especially to Fu-Sheng Yu and Yan-Bing Wei for their inspiring discussions on various experimental and theoretical aspects. Special thanks go to Long Chen and Yan-Qing Ma for providing the numerical NNLO corrections to the partonic decay widths and the corresponding electron energy moments. This work is supported in part by Natural Key R$\&$D Program of China under Contract No. 2023YFA1609400; National Natural Science Foundation of China (NSFC) under Contract No.~12105127.

}





\bibliographystyle{iopart-num}
\bibliography{references}

@article{Shao:2025vhe,
    author = "Shao, Kang-Kang and Huang, Chun and Qin, Qin",
    title = "{Data determination of HQET parameters in inclusive charm decays}",
    eprint = "2502.05901",
    archivePrefix = "arXiv",
    primaryClass = "hep-ph",
    doi = "10.1140/epjc/s10052-025-14691-z",
    journal = "Eur. Phys. J. C",
    volume = "85",
    number = "9",
    pages = "1011",
    year = "2025"
}

@article{Bernlochner:2024vhg,
    author = "Bernlochner, Florian and Gilman, Alex and Malde, Sneha and Prim, Markus and Vos, K. Keri and Wilkinson, Guy",
    title = "{Charming Darwin: the evolution of QCD parameters across different species}",
    eprint = "2408.10063",
    archivePrefix = "arXiv",
    primaryClass = "hep-ex",
    doi = "10.1007/JHEP05(2025)061",
    journal = "JHEP",
    volume = "05",
    pages = "061",
    year = "2025"
}

@article{Shao:2025qwp,
    author = "Shao, Kang-Kang and Feng, Hai-Long and Liu, Xue-Yin and Qin, Qin and Sun, Liang and Yu, Fu-Sheng",
    title = "{First determination of $V_{cs,cd}$ from inclusive $D$ meson decays}",
    eprint = "2509.11404",
    archivePrefix = "arXiv",
    primaryClass = "hep-ph",
    month = "9",
    year = "2025"
}

@article{Gratrex:2022xpm,
    author = "Gratrex, James and Meli{\'c}, Bla{\v{z}}enka and Ni{\v{s}}and{\v{z}}i{\'c}, Ivan",
    title = "{Lifetimes of singly charmed hadrons}",
    eprint = "2204.11935",
    archivePrefix = "arXiv",
    primaryClass = "hep-ph",
    reportNumber = "RBI-ThPhys-2022-8",
    doi = "10.1007/JHEP07(2022)058",
    journal = "JHEP",
    volume = "07",
    pages = "058",
    year = "2022"
}

@article{LHCb:2019ldj,
    author = "Aaij, Roel and others",
    collaboration = "LHCb",
    title = "{Precision measurement of the $\Lambda_c^+$, $\Xi_c^+$ and $\Xi_c^0$ baryon lifetimes}",
    eprint = "1906.08350",
    archivePrefix = "arXiv",
    primaryClass = "hep-ex",
    reportNumber = "LHCb-PAPER-2019-008, CERN-EP-2019-122",
    doi = "10.1103/PhysRevD.100.032001",
    journal = "Phys. Rev. D",
    volume = "100",
    number = "3",
    pages = "032001",
    year = "2019"
}

@article{PhysRevD.107.052005,
  title = {Improved measurement of the absolute branching fraction of inclusive semileptonic ${\mathrm{\ensuremath{\Lambda}}}_{c}^{+}$ decay},
  author = {M. Ablikim and others},
  collaboration = {BESIII Collaboration},
  journal = {Phys. Rev. D},
  volume = {107},
  issue = {5},
  pages = {052005},
  numpages = {12},
  year = {2023},
  month = {Mar},
  publisher = {American Physical Society},
  doi = {10.1103/PhysRevD.107.052005},
  url = {https://link.aps.org/doi/10.1103/PhysRevD.107.052005}
}

@misc{dagostini2010,
      title={Improved iterative Bayesian unfolding}, 
      author={G. D'Agostini},
      year={2010},
      eprint={1010.0632},
      archivePrefix={arXiv},
      primaryClass={physics.data-an},
      url={https://arxiv.org/abs/1010.0632}, 
}

@article{LHCb:2021vll,
    author = "Aaij, Roel and others",
    collaboration = "LHCb",
    title = "{Measurement of the lifetimes of promptly produced $\Omega^{0}_{c}$ and $\Xi^{0}_{c}$ baryons}",
    eprint = "2109.01334",
    archivePrefix = "arXiv",
    primaryClass = "hep-ex",
    reportNumber = "LHCb-PAPER-2021-021, CERN-EP-2021-167",
    doi = "10.1016/j.scib.2021.11.022",
    journal = "Sci. Bull.",
    volume = "67",
    number = "5",
    pages = "479--487",
    year = "2022"
}

@article{LHCb:2018nfa,
    author = "Aaij, Roel and others",
    collaboration = "LHCb",
    title = "{Measurement of the $\Omega_c^0$ baryon lifetime}",
    eprint = "1807.02024",
    archivePrefix = "arXiv",
    primaryClass = "hep-ex",
    reportNumber = "LHCb-PAPER-2018-028, CERN-EP-2018-175, LHCB-PAPER-2018-028",
    doi = "10.1103/PhysRevLett.121.092003",
    journal = "Phys. Rev. Lett.",
    volume = "121",
    number = "9",
    pages = "092003",
    year = "2018"
}

@article{Cheng:2023jpz,
    author = "Cheng, Hai-Yang and Liu, Chia-Wei",
    title = "{Study of singly heavy baryon lifetimes}",
    eprint = "2305.00665",
    archivePrefix = "arXiv",
    primaryClass = "hep-ph",
    doi = "10.1007/JHEP07(2023)114",
    journal = "JHEP",
    volume = "07",
    pages = "114",
    year = "2023"
}

@article{Cheng:2021qpd,
    author = "Cheng, Hai-Yang",
    title = "{Charmed baryon physics circa 2021}",
    eprint = "2109.01216",
    archivePrefix = "arXiv",
    primaryClass = "hep-ph",
    doi = "10.1016/j.cjph.2022.06.021",
    journal = "Chin. J. Phys.",
    volume = "78",
    pages = "324--362",
    year = "2022"
}

@article{Fael:2019umf,
    author = "Fael, Matteo and Mannel, Thomas and Vos, K. Keri",
    title = "{The Heavy Quark Expansion for Inclusive Semileptonic Charm Decays Revisited}",
    eprint = "1910.05234",
    archivePrefix = "arXiv",
    primaryClass = "hep-ph",
    reportNumber = "SI-HEP-2019-07, P3H-19-34, TPP19-032, TUM-HEP-1228/19, INT-PUB-19-049, TUM-HEP-1228/19,
  INT-PUB-19-049",
    doi = "10.1007/JHEP12(2019)067",
    journal = "JHEP",
    volume = "12",
    pages = "067",
    year = "2019"
}

@article{Hoang:1998ng,
    author = "Hoang, Andre H. and Ligeti, Zoltan and Manohar, Aneesh V.",
    title = "{B decay and the Upsilon mass}",
    eprint = "hep-ph/9809423",
    archivePrefix = "arXiv",
    reportNumber = "UCSD-PTH-98-31",
    doi = "10.1103/PhysRevLett.82.277",
    journal = "Phys. Rev. Lett.",
    volume = "82",
    pages = "277--280",
    year = "1999"
}

@article{Hoang:1998hm,
    author = "Hoang, Andre H. and Ligeti, Zoltan and Manohar, Aneesh V.",
    title = "{B decays in the upsilon expansion}",
    eprint = "hep-ph/9811239",
    archivePrefix = "arXiv",
    reportNumber = "UCSD-PTH-98-32, CERN-TH-98-334, FERMILAB-PUB-98-351-T",
    doi = "10.1103/PhysRevD.59.074017",
    journal = "Phys. Rev. D",
    volume = "59",
    pages = "074017",
    year = "1999"
}

@article{Hoang:1999zc,
    author = "Hoang, A. H. and Teubner, T.",
    title = "{Top quark pair production close to threshold: Top mass, width and momentum distribution}",
    eprint = "hep-ph/9904468",
    archivePrefix = "arXiv",
    reportNumber = "CERN-TH-99-59, DESY-99-047",
    doi = "10.1103/PhysRevD.60.114027",
    journal = "Phys. Rev. D",
    volume = "60",
    pages = "114027",
    year = "1999"
}

@article{FlavourLatticeAveragingGroup:2019iem,
    author = "Aoki, S. and others",
    collaboration = "Flavour Lattice Averaging Group",
    title = "{FLAG Review 2019: Flavour Lattice Averaging Group (FLAG)}",
    eprint = "1902.08191",
    archivePrefix = "arXiv",
    primaryClass = "hep-lat",
    reportNumber = "FERMILAB-PUB-19-077-T",
    doi = "10.1140/epjc/s10052-019-7354-7",
    journal = "Eur. Phys. J. C",
    volume = "80",
    number = "2",
    pages = "113",
    year = "2020"
}

@article{ParticleDataGroup:2024cfk,
    author = "Navas, S. and others",
    collaboration = "Particle Data Group",
    title = "{Review of particle physics}",
    doi = "10.1103/PhysRevD.110.030001",
    journal = "Phys. Rev. D",
    volume = "110",
    number = "3",
    pages = "030001",
    year = "2024"
}

@article{King:2021xqp,
    author = "King, Daniel and Lenz, Alexander and Piscopo, Maria Laura and Rauh, Thomas and Rusov, Aleksey V. and Vlahos, Christos",
    title = "{Revisiting inclusive decay widths of charmed mesons}",
    eprint = "2109.13219",
    archivePrefix = "arXiv",
    primaryClass = "hep-ph",
    reportNumber = "SI-HEP-2021-23, SFB-257-P3H-21-058, IPPP/21/22",
    doi = "10.1007/JHEP08(2022)241",
    journal = "JHEP",
    volume = "08",
    pages = "241",
    year = "2022"
}

@article{Herren:2017osy,
    author = "Herren, Florian and Steinhauser, Matthias",
    title = "{Version 3 of RunDec and CRunDec}",
    eprint = "1703.03751",
    archivePrefix = "arXiv",
    primaryClass = "hep-ph",
    reportNumber = "TTP17-011",
    doi = "10.1016/j.cpc.2017.11.014",
    journal = "Comput. Phys. Commun.",
    volume = "224",
    pages = "333--345",
    year = "2018"
}

@article{Friday:2025gpj,
    author = "Friday, David and Gersabeck, Evelina and Lenz, Alexander and Piscopo, Maria Laura",
    title = "{Charm physics}",
    eprint = "2506.15584",
    archivePrefix = "arXiv",
    primaryClass = "hep-ph",
    reportNumber = "Nikhef 2025-009, SI-HEP-2025-14, P3H-25-040",
    month = "6",
    year = "2025"
}

@article{Egner:2024lay,
    author = {Egner, Manuel and Fael, Matteo and Lenz, Alexander and Piscopo, Maria Laura and Rusov, Aleksey V. and Sch{\"o}nwald, Kay and Steinhauser, Matthias},
    title = "{Total decay rates of B mesons at NNLO-QCD}",
    eprint = "2412.14035",
    archivePrefix = "arXiv",
    primaryClass = "hep-ph",
    reportNumber = "TUM-HEP-1545/24, P3H-24-101, SI-HEP-2024-31, TTP24-046, Nikhef
  2024-019, ZU-TH 67/24",
    doi = "10.1007/JHEP04(2025)106",
    journal = "JHEP",
    volume = "04",
    pages = "106",
    year = "2025"
}

@article{Black:2024bus,
    author = {Black, Matthew and Lang, Martin and Lenz, Alexander and W{\"u}thrich, Zachary},
    title = "{HQET sum rules for matrix elements of dimension-six four-quark operators for meson lifetimes within and beyond the Standard Model}",
    eprint = "2412.13270",
    archivePrefix = "arXiv",
    primaryClass = "hep-ph",
    reportNumber = "P3H-24-098, SI-HEP-2024-29",
    doi = "10.1007/JHEP04(2025)081",
    journal = "JHEP",
    volume = "04",
    pages = "081",
    year = "2025"
}

@article{Piscopo:2024wpd,
    author = "Piscopo, Maria Laura and Lenz, Alexander and Rusov, Aleksey V.",
    title = "{Towards a SM prediction for CP violation in charm}",
    eprint = "2403.02267",
    archivePrefix = "arXiv",
    primaryClass = "hep-ph",
    doi = "10.22323/1.443.0028",
    journal = "PoS",
    volume = "BEAUTY2023",
    pages = "028",
    year = "2024"
}

@article{King:2021jsq,
    author = "King, Daniel and Lenz, Alexander and Rauh, Thomas",
    title = "{SU(3) breaking effects in B and D meson lifetimes}",
    eprint = "2112.03691",
    archivePrefix = "arXiv",
    primaryClass = "hep-ph",
    reportNumber = "IPPP/21/38, SI-HEP-2021-026, SFB-257-P3H-21-071",
    doi = "10.1007/JHEP06(2022)134",
    journal = "JHEP",
    volume = "06",
    pages = "134",
    year = "2022"
}

@article{Kirk:2017juj,
    author = "Kirk, M. and Lenz, A. and Rauh, T.",
    title = "{Dimension-six matrix elements for meson mixing and lifetimes from sum rules}",
    eprint = "1711.02100",
    archivePrefix = "arXiv",
    primaryClass = "hep-ph",
    reportNumber = "IPPP/17/65, IPPP-17-65",
    doi = "10.1007/JHEP12(2017)068",
    journal = "JHEP",
    volume = "12",
    pages = "068",
    year = "2017",
    note = "[Erratum: JHEP 06, 162 (2020)]"
}

@article{Dulibic:2023jeu,
    author = "Dulibi{\'c}, Lovro and Gratrex, James and Meli{\'c}, Bla{\v{z}}enka and Ni{\v{s}}and{\v{z}}i{\'c}, Ivan",
    title = "{Revisiting lifetimes of doubly charmed baryons}",
    eprint = "2305.02243",
    archivePrefix = "arXiv",
    primaryClass = "hep-ph",
    reportNumber = "RBI-ThPhys-2023-9",
    doi = "10.1007/JHEP07(2023)061",
    journal = "JHEP",
    volume = "07",
    pages = "061",
    year = "2023"
}

@article{Boushmelev:2023kmf,
    author = "Boushmelev, Anastasia and Mannel, Thomas and Vos, K. Keri",
    title = "{Alternative treatment of the quark mass in the heavy quark expansion}",
    eprint = "2301.05607",
    archivePrefix = "arXiv",
    primaryClass = "hep-ph",
    reportNumber = "SI-HEP-2023-01, NIKHEF-2023-001, SI-HEP-2023-01 NIKHEF-2023-001",
    doi = "10.1007/JHEP07(2023)175",
    journal = "JHEP",
    volume = "07",
    pages = "175",
    year = "2023"
}

@article{Gray:1990yh,
    author = "Gray, N. and Broadhurst, David J. and Grafe, W. and Schilcher, K.",
    title = "{Three Loop Relation of Quark (Modified) Ms and Pole Masses}",
    reportNumber = "OUT-4102-28",
    doi = "10.1007/BF01614703",
    journal = "Z. Phys. C",
    volume = "48",
    pages = "673--680",
    year = "1990"
}

@article{Broadhurst:1991fy,
    author = "Broadhurst, David J. and Gray, N. and Schilcher, K.",
    title = "{Gauge invariant on-shell Z(2) in QED, QCD and the effective field theory of a static quark}",
    reportNumber = "OUT-4102-29",
    doi = "10.1007/BF01412333",
    journal = "Z. Phys. C",
    volume = "52",
    pages = "111--122",
    year = "1991"
}

@article{Fleischer:1998dw,
    author = "Fleischer, J. and Jegerlehner, F. and Tarasov, O. V. and Veretin, O. L.",
    title = "{Two loop QCD corrections of the massive fermion propagator}",
    eprint = "hep-ph/9803493",
    archivePrefix = "arXiv",
    reportNumber = "DESY-98-026",
    doi = "10.1016/S0550-3213(98)00705-6",
    journal = "Nucl. Phys. B",
    volume = "539",
    pages = "671--690",
    year = "1999",
    note = "[Erratum: Nucl.Phys.B 571, 511--512 (2000)]"
}

@article{Melnikov:2000qh,
    author = "Melnikov, Kirill and Ritbergen, Timo van",
    title = "{The Three loop relation between the MS-bar and the pole quark masses}",
    eprint = "hep-ph/9912391",
    archivePrefix = "arXiv",
    reportNumber = "SLAC-PUB-8321, TTP-99-51",
    doi = "10.1016/S0370-2693(00)00507-4",
    journal = "Phys. Lett. B",
    volume = "482",
    pages = "99--108",
    year = "2000"
}

\end{document}